# Low Temperature Shear Modulus Changes in Solid $^4$He and Connection to Supersolidity

James Day and John Beamish

*Department of Physics, University of Alberta, Edmonton, Alberta CANADA T6G 2G7*

**Superfluidity, liquid flow without friction, is familiar in helium. The first evidence for "supersolidity", its analogue in quantum solids, came from recent torsional oscillator (TO) measurements[1,2] involving $^4$He. At temperatures below 200 mK, TO frequencies increased, suggesting that some of the solid decoupled from the oscillator. This behavior has been replicated by several groups[3,4,5,6,7], but solid $^4$He does not respond to pressure differences[8] and persistent currents and other signatures of superflow have not been seen. Both experiments and theory[9,10,11,12,13,14] indicate that defects are involved. These should also affect the solid's mechanical behavior and so we have measured the shear modulus of solid $^4$He at low frequencies and strains. We observe large increases below 200 mK, with the same dependence on measurement amplitude, $^3$He impurity concentration and annealing as the decoupling seen in TO experiments. This unusual elastic behavior is explained in terms of a dislocation network which is pinned by $^3$He at the lowest temperatures but becomes mobile above 100 mK. The frequency changes in TO experiments appear to be related to the motion of these dislocations, perhaps by disrupting a possible supersolid state.**

Although the amount of helium which decouples in different TOs varies widely, the measurements have many common features. Decoupling occurs below about 200 mK, with a gradual onset accompanied by a dissipation peak at lower temperatures. It decreases at large oscillation amplitudes, which is interpreted in terms of a superflow critical velocity ($v_c \approx 10$ μm/s). The magnitude of the decoupling is frequency independent, although its onset shifts with frequency[7]. Its amplitude dependence appears to scale with velocity but depends on the oscillation amplitude during cooling and is hysteretic. A crucial feature of the decoupling is its sensitivity to $^3$He. Most measurements used commercial $^4$He gas (with $^3$He concentration $x_3 \sim 0.3$ ppm), but experiments[15,16] with isotopically pure $^4$He (1 ppb $^3$He) show a sharper onset at a lower temperature, around 75 mK. Decoupling is usually larger in narrow annuli than in open cylinders[6] but begins at similar temperatures. Its magnitude also depends on how the solid helium was grown and annealed, indicating that defects are important. Most samples were grown at constant volume under "blocked capillary" conditions – a procedure which involves substantial plastic deformation and is expected to produce a polycrystalline solid with many defects. Theoretical work also suggests that supersolidity does not occur in a perfect crystal[9,10] and that grain boundaries[11], glassy regions[12] or dislocations[13,14] are involved. Superflow associated with grain boundaries has been seen in solid $^4$He coexisting with liquid[17], but solidification at constant pressure (producing single crystals with fewer defects) still gives significant TO decoupling[16]. A recently reported heat capacity peak[18] supports the existence of a new phase where decoupling occurs.

We have made a detailed study of the elastic properties of solid $^4$He. This required a new method to measure the shear modulus $\mu$ at extremely low frequencies and amplitudes, which proved crucial. Embedding piezoelectric transducers in the helium (see Supplementary Figure 1 and Supplementary Methods) allowed us to measure $\mu$ directly at strains (stresses) as low as $\varepsilon = 2.2 \times 10^{-9}$ ($\sigma = 0.03$ Pa). This is two to three orders of magnitude lower than in previous torsional[19], internal friction[20] and ultrasonic[21,22] measurements. We measured $\mu$ at frequencies down to 20 Hz, far lower than other experiments. We could also excite and detect acoustic modes of solid helium outside the gap, surrounding the transducers. The first such resonance was near 8000 Hz and had a quality factor $Q \sim 2000$ at our lowest temperature.

Our essential result is the observation of a large anomalous increase in $\mu$ with the same temperature dependence as the decoupling in TOs. We confirmed this effect by simultaneously measuring the frequency $f_r$ and damping $1/Q$ of an acoustic resonance in the cell. A comprehensive study of the anomaly's dependence on frequency, amplitude and $^3$He concentration showed all the same features as the TO decoupling and provided new information about the effects of annealing and stressing the solid. This is the first clear observation of directly related phenomena in other properties of solid helium. Our measurements are consistent with the known behavior of dislocations in solid helium and lead to a picture of a dislocation network pinned at low temperatures by bound $^3$He impurities. Above about 100 mK, $^3$He atoms unbind and allow dislocations to move in response to stress thus weakening the crystal.

Figure 1 shows the temperature dependence of the shear modulus of hcp $^4$He at pressures of 29.3 and 33.3 bar (with melting temperatures $T_m$ = 1.77 and 1.86 K - the first sample passed along the bcc/hcp line during growth). The bottom curve shows $\mu$ at 2000 Hz in the first sample. Below 200 mK, it increases by about 11% ($\Delta\mu$ ~ 16 bar) - this anomalous stiffening is our central result. The middle curves show this behavior at three frequencies (2000, 200 and 20 Hz) in the second sample. The magnitude of the modulus increase, $\Delta\mu$, is similar (~ 8%) and is nearly independent of frequency, although the transition is sharper at low frequency. The pressure in the cell is constant within 0.2 mbar in this temperature range, which rules out local density changes (e.g. freezing of small liquid regions) as the cause of the $\mu$ increase and implies that the bulk modulus does not have a similar anomaly. The upper curve shows a typical NCRI fraction from a TO measurement[2] (at a frequency of 910 Hz). The temperature dependence (onset and shape) is essentially the same as that of shear modulus anomaly. We observed variations in $\Delta\mu$ of up to a factor of 2 (over a total of 8 samples), similar to the range of NCRI seen in a single TO.

Figure 2 shows $\mu$ for the 33.3 bar sample at different strains (calculated from drive voltages). The anomaly $\Delta\mu$ is independent of drive amplitude for strains up to $2.2 \times 10^{-8}$ then begins to decrease. We observed almost identical behavior at 200 Hz - the amplitude dependence began at the same drive level, indicating that it scales with stress or strain and not with velocity. The corresponding velocities ($\sim$ 50 nm/s for $\varepsilon = 2.2 \times 10^{-8}$ at 2000 Hz) are much smaller than the critical velocity $v_c \approx 10$ $\mu$m/s inferred from TO measurements. However, the stress levels (0.3 Pa) are comparable to inertial stresses in torsional oscillators (e.g., $\sigma_t \approx 0.15$ Pa at the highest velocity, 520 $\mu$m/s, in the TO of ref. 2). The behavior is reversible at temperatures above the anomaly and at low amplitudes. When a sample is cooled at high amplitude and the drive is then reduced at low temperature, $\mu$ increases (as expected from Fig. 2). However, $\mu$ does not decrease when the amplitude is then raised again. Similar hysteretic behavior is seen in TO decoupling[7]. Previous torsional measurements[19] at comparable strains ($\varepsilon = 10^{-7}$) showed no change in $\mu$ between 0.5 K and 17 mK, an apparent discrepancy with our results.

Figure 3 shows the behavior of the acoustic resonance near 8 kHz, for the same sample. Comparing the temperature dependence of the resonance frequency $f_r$ to that of the shear modulus $\mu$, it is clear that the two measurements probe the same elastic changes. The changes in $f_r$ are about half as large as for $\mu$, as expected since $f_r$ varies with sound speed, i.e. with the square root of elastic moduli. The corresponding dissipation $1/Q$ is largest near 150 mK, where $f_r$ is changing rapidly (the dissipation peak occurs at lower temperatures in most TO measurements, but this may be due to their lower frequencies, 185 to 1500 Hz). In a simple oscillator the maximum dissipation $\Delta(1/Q)$ should equal the frequency shift $\Delta f_r/f_r$, but in our case it is a factor of about 3 smaller. Similar differences in TO measurements have been ascribed[23] to sample inhomogeneity.

A striking feature of TO experiments is their sensitivity to $^3$He. We grew samples from isotopically pure $^4$He (1 ppb $^3$He - the same gas used in TO measurements[15]) and from intermediate concentrations made by mixing with commercial $^4$He (0.3 ppm $^3$He). We compare their behavior in Fig. 4. Changes have been scaled by $\Delta\mu$ at 18 mK to compare temperature dependences. The anomaly shifts to lower temperatures as the $^3$He concentration decreases. We also show similarly scaled decoupling data from TO experiments[2,16] on 1 ppb and 0.3 ppm $^3$He samples: the onset temperatures and shapes of the curves agree very well (within the sample to sample variations in TO measurements).

To understand the role of defects, we annealed the 33.3 bar sample at 160 mK below its melting point $T_m$ for 15 hours. This reduced $\Delta\mu$ from 9.8 to 7.7%, but it was the high temperature behavior which changed. The low temperature values of $\mu$ and $f_r$ were almost unaffected (e.g., at 18 mK, $f_r$ increased by only 0.1%) and appear to reflect the intrinsic shear modulus, so the effect of annealing is to change $\mu$ at higher temperatures. We also applied large acoustic stresses ($\geq$ 700 Pa) to the annealed crystal in an attempt to create additional defects. Again, the values of $\mu$ and $f_r$ changed at high temperature but not at low temperature. Warming above ~ 0.6 K undoes these effects, indicating that defects introduced by stressing the crystal are only stable at low temperatures. Ultrasonic measurements[24] on bcc and hcp $^3$He showed similar effects of large stresses.

The modulus changes we see are much larger than expected in defect-free crystals. Dislocations (1-dimensional defects created during crystal growth or deformation) can dramatically affect elastic properties and lead to unusual behavior in quantum crystals[25]. They form 3-dimensional networks, pinned where they intersect, and characterized by their density $\Lambda$ (total length per unit volume), Burgers vector b, and length between intersections $L_N$. They can also be pinned, less strongly, by impurities with the pinning length $L_P$ determined by the impurity concentration x, binding energy $E_B$, and the temperature.

Dislocations move in response to stress, producing a strain[26]. At low frequencies this reduces the shear modulus by $\Delta\mu/\mu = -CR\Lambda L^2$ (see Supplementary Discussion) where C is a constant ($\leq 0.5$) which depends on the distribution of lengths and R is an orientation factor (~ 0.2). In annealed crystals with well-defined networks, $\Lambda L_N^2$ is a geometric constant (e.g., 3 for a cubic network), so dislocations can produce a frequency-independent reduction of the shear modulus as large as 30%. Impurities can immobilize dislocations and restore the crystal's intrinsic modulus. This pinning becomes important when $L_P$ is comparable to $L_N$, which occurs at a temperature $T_P$ which decreases with impurity concentration: $T_P \sim -(E_B/k_B) \cdot (\ln\{x \cdot L_N/a\})^{-1}$, where a is the interatomic spacing ($\approx 0.35$ nm).

Our results are consistent with a picture of a network of dislocations pinned at low temperature by $^3$He impurities, using dislocation parameters determined in earlier experiments on hcp $^4$He. Low frequency torsional measurements[19] on helium gave a $^3$He binding energy $E_B/k_B \approx 0.6$ K and $R\Lambda L^2 \approx 1$. Ultrasonic measurements[21,22] on single crystals gave $L_N \approx 5$ μm. Using these values, $T_P \approx 110$ mK for $x_3 = 0.3$ ppm, decreasing to 54 mK for $x_3 = 1$ ppb, close to the temperatures where μ increases. Large stresses can tear dislocations away from $^3$He pinning sites and reduce the shear modulus[27]. The critical stress for this breakaway can be estimated[27] as 4 Pa for $L_P \approx 5$ μm. This corresponds to $\varepsilon \sim 3 \times 10^{-7}$, a strain where we see strong amplitude dependence. Annealing should reduce the dislocation density $\Lambda$ (and large stresses can increase $\Lambda$), but $\Delta\mu$ depends on the combination $\Lambda L^2$, which doesn't necessarily decrease when dislocations disappear. The low temperature behavior is unaffected by annealing or stressing, as expected since dislocations are then pinned by impurities and don't affect μ. When the temperature is raised, $^3$He impurities "boil off" the dislocations, allowing them to move and reducing μ.

Is our shear modulus anomaly directly related to the frequency shifts and dissipation in TO experiments? Although they measure very different properties (shear modulus and sound speed versus moment of inertia and density), the two sets of measurements share all essential features. The anomalous behavior has the same temperature dependence and the transitions are accompanied by similar dissipation peaks. Both are strongly amplitude dependent (starting at comparable stress levels) and have similar amplitude-dependent hysteresis at low temperatures. In both types of measurements, the anomaly's magnitude is frequency independent, but its onset is broadened and shifts to higher temperature with increasing frequency. Minute $^3$He concentrations have the same dramatic effect on the onset temperature and annealing changes the size of both anomalies. Given these remarkable similarities, the two sets of effects must be closely related. The obvious question is "how?"

One possibility is that the modulus increase stiffens the TO, increasing its frequency and mimicking mass decoupling. Interpreting a TO frequency as a direct measure of mass assumes that the oscillator head is infinitely stiff and that the solid helium moves rigidly with its walls, neither of which is exactly true. However, an increase in the helium's shear modulus would improve its coupling to the TO and thus decrease its frequency (i.e., the opposite of the observed behavior). An increase in μ could raise an oscillator's frequency by increasing the stiffness of its head. However, the head itself is much stiffer than the torsion rod and the helium's contribution to the head's stiffness is small (it has moduli about 3000 times smaller than copper), so modulus changes in the helium should produce very small frequency changes. Estimates for typical oscillators indicate that this effect is several orders of magnitude too small to explain the observed decoupling. Also, blocking the flow path in a TO annulus would barely change the helium's contribution to its stiffness, yet nearly eliminated the decoupling[2]. Our observations do not provide an obvious mechanical, non-supersolid explanation of the frequency changes in TOs.

Our µ anomaly and the decoupling seen in TO measurements could both be fundamental properties of a supersolid state[28], in which case it would be natural for them to have a common dependence on temperature, $^3$He, etc. This could be the case if, for example, supersolidity occurs along the cores[13] or strain fields[14] of dislocation networks. Alternatively, the µ anomaly could, as we propose, be due to dislocations becoming mobile, which in turn could affect a supersolid response. For example, vortices[29] could be pinned by stationary dislocations but could introduce dissipation and destroy the supersolidity when dislocations begin to move above 100 mK. The decoupling seen in porous media[1] remains a puzzle, since it is hard to imagine dislocations existing, let alone moving, in the 7 nm pores of vycor. The precise connection between our elastic measurements and decoupling of solid helium from TOs is not certain, but it is clear that the two are closely related and that models of supersolidity should consider the effects of moving dislocations.

**References**

1.      Kim, E. and Chan, M.H.W. Probable observation of a supersolid helium phase. *Nature* **427**, 225 (2004).

2.      Kim, E. and Chan, M.H.W. Observation of superflow in solid helium. *Science* **305**, 1941 (2004).

3.      Kondo, M., Takada, S., Shibayama, Y. and Shirahama, K. Observation of non-classical rotational inertia in bulk solid $^4$He. *J. Low Temp. Phys.* **148**, 695 (2007).

4.      Penzev, A., Yasuta, Y. and Kubota, M. Annealing effect for supersolid fraction in $^4$He. *J. Low Temp. Phys.* **148**, 677 (2007).

5.      Rittner, A.S.C. and Reppy, J.D. Observation of classical rotational inertia and nonclassical supersolid signals in solid $^4$He below 250 mK. *Phys. Rev. Lett.* **97**, 165301 (2006).

6.       Rittner, A.S.C. and Reppy, J.D. Disorder and the supersolid state of solid $^4$He. *Phys. Rev. Lett.* **98**, 175302 (2007).

7.      Aoki,Y., Graves, J.C. and Kojima, H. Oscillation frequency dependence of nonclassical rotation inertia of solid $^4$He. *Phys. Rev. Lett.* **99**, 015301 (2007).

8.      Day, J. and Beamish, J. Pressure-driven flow of solid helium. *Phys. Rev. Lett*. **96** 105304 (2006).

9.      Ceperley, D.M. and Bernu, B. Ring exchanges and the supersolid phase of $^4$He. *Phys. Rev. Lett.* **93**, 155303 (2004).

10.     Prokof'ev, N. and Svistunov, B. Supersolid state of matter. *Phys. Rev. Lett.* **94**, 155302 (2005).

11.     Pollet, L. et al. Superfluidity of grain boundaries in solid $^4$He. *Phys. Rev. Lett.* **98**, 135301 (2007).

12.     Boninsegni, M., Prokof'ev, N., and Svistunov, B. Superglass phase of $^4$He. *Phys. Rev. Lett.* **96**, 105301 (2006).

13.     Boninsegni, M. et al. Luttinger liquid in the core of a screw dislocation in helium-4. *Phys. Rev. Lett.* **99**, 035301 (2007).

14.     Toner, J. Quenched disorder enhanced supersolid ordering. *arXiv*:0707.3842 (2007).

15.     Kim, E., Xia, J. S., West, J.T., Lin, X, and Chan, M.H.W. Effect of $^3$He impurity on the supersolid transition of $^4$He. *Bull. Am. Phys. Soc.* **52**, 610 (2007).

16.     Clark, A.C., West, J.T., and Chan, M.H.W. Non-classical rotational inertia in helium crystals. *arXiv:cond-mat*/0706.0906 (2007).

17.     Sasaki, S., Ishiguro, R., Caupin, F., Maris, H.J. and Balibar, S. Superfluidity of grain boundaries and supersolid behavior. *Science* **313**, 1098 (2006).

18.     Lin, X, Clark, A.C. and Chan, M.H.W. Excess specific heat peak of solid $^4$He. *Bull. Am. Phys. Soc.* **52**, 610 (2007).

19.     Paalanen, M.A., Bishop, D.J., and Dail, H.W. Dislocation motion in hcp $^4$He. *Phys. Rev. Lett.* **46**, 664 (1981).

20.     Tsymbalenko, V.L. Measurement of internal friction in solid $^4$He. *Sov. Phys. JETP* 47, 787 (1978).

21.     Iwasa, I., Araki, K. and Suzuki, H. Temperature and frequency dependence of the sound velocity in hcp $^4$He crystals. *J. Phys. Soc. Japan* **46**, 1119 (1979);

22.     Beamish, J.R. and Franck, J.P. Sound propagation at frequencies from 3 to 21 MHz in hcp and bcc $^3$He and its interaction with dislocations. *Phys. Rev*. **B26**, 6104 (1982).

23.     Huse, D.A. and Khandker, Z.U. Dissipation peak as an indicator of sample inhomogeneity in solid $^4$He oscillator experiments. *condmat*/0702243 (2007).

24. Beamish, J.R. and Franck, J.P. Pinning of dislocations in hcp and bcc $^3$He by stress waves and by $^4$He impurities. *Phys. Rev.* **B28**, 1419 (1983).

25. De Gennes, P.-G. Quantum dynamics of a single dislocation. *C. R. Physique* **7**, 561 (2006).

26. Granato, A. and Lucke, K. Theory of mechanical damping due to dislocations. *J. Appl. Phys.* **27**, 583 (1956).

27. Iwasa, I. and Suzuki, H. Sound velocity and attenuation in hcp $^4$He crystals containing $^3$He impurities. *J. Phys. Soc. Japan* **49**, 1722 (1980).

28. Dorsey, A.T., Goldbart, P.M. and Toner, J. Squeezing superfluid from a stone: coupling superfluidity and elasticity in a supersolid. *Phys. Rev. Lett.* 96, 055301 (2006).

29. Anderson, P.W. Two new vortex fluids. *Nature Physics* **3**, 160 (2007).

We would like to thank the Natural Sciences and Engineering Research Council of Canada and the University of Alberta for support of this research and M.H.W. Chan for providing the TO data of Figs. 1 and 4.

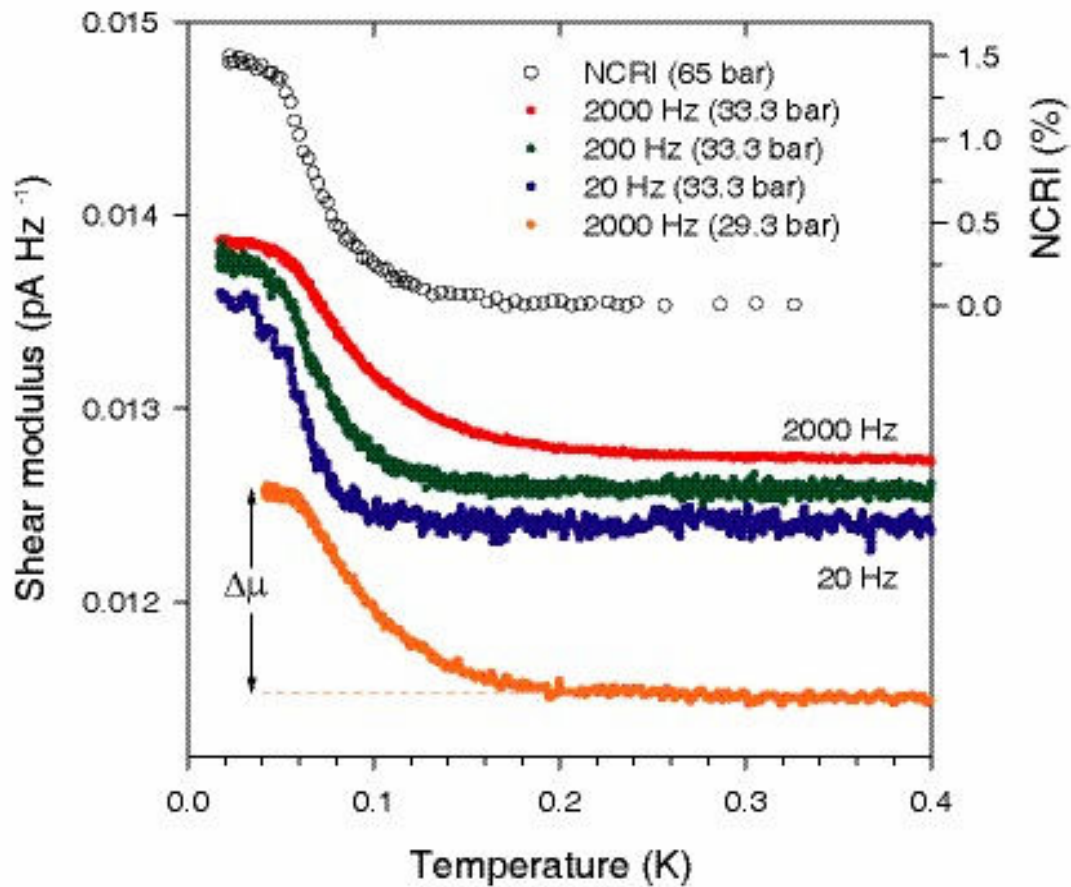

Figure 1: Shear modulus (l/f) of solid $^4$He at $\varepsilon = 2.2 \times 10^{-8}$ as a function of temperature. Data have been offset for clarity. Bottom (orange) curve: $\mu$ at 2000 Hz in a 29.3 bar sample. Middle three curves: $\mu$ at 2000 Hz (red), 200 Hz (green), and 20 Hz (blue) in a 33.3 bar sample. Top curve (open circles, right axes): typical NCRI fraction from a torsional oscillator measurement[1] in a 65 bar sample.

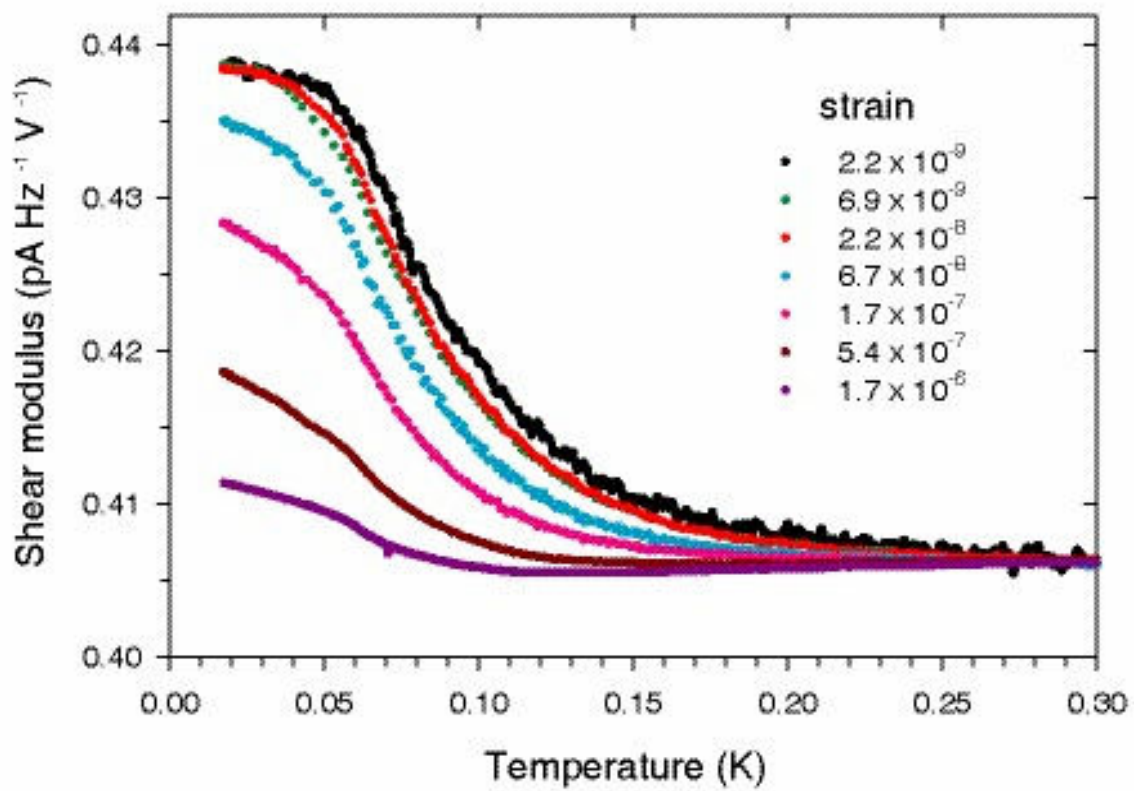

Figure 2: Shear modulus (I/fV) at 2000 Hz as a function of peak strain amplitude ($\varepsilon = 2.2 \times 10^{-9}$ to $1.7 \times 10^{-6}$, calculated from transducer drive voltage) in the 33.3 bar sample. Data for different strains have been shifted to coincide at 300 mK. All data were taken during cooling.

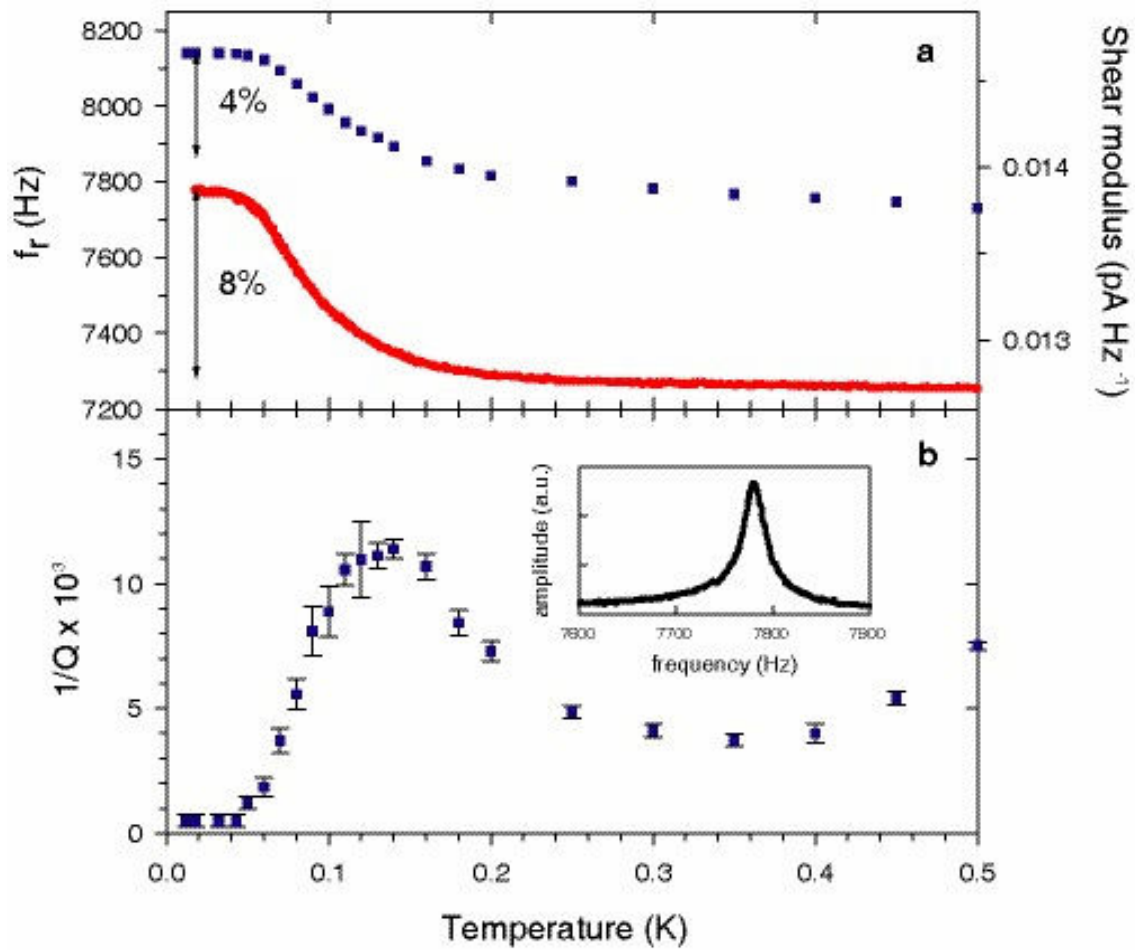

Figure 3: Temperature dependence of acoustic resonance peak frequency $f_r$ and dissipation $1/Q$ for the 33.3 bar sample. Fig 4a. (upper panel): resonance frequency (upper blue points, left axis) and, for comparison, shear modulus at 2000 Hz (lower red curve, right axis). Vertical arrows are for scale, showing 4% and 8% changes in $f_r$ and $\mu$, respectively. Fig. 4b. (lower panel): dissipation (calculated from the peaks' full widths at half maximum) corresponding to resonance data in Fig. 4a. Inset: typical resonance peak at 300 mK with a Q ~ 250.

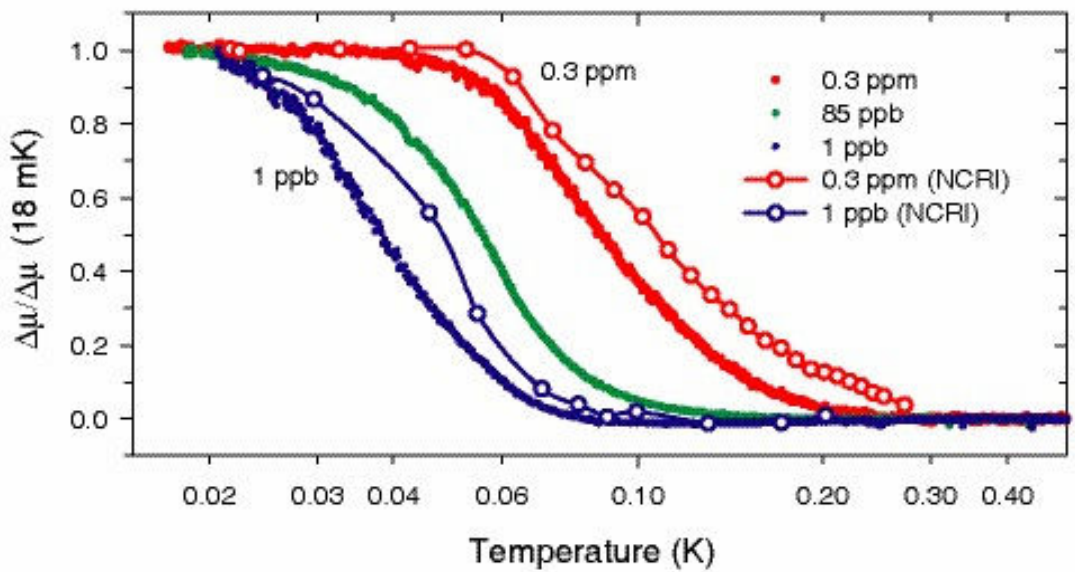

Figure 4: Shear modulus anomaly in solid $^4$He for the $^3$He impurity concentrations: 1 ppb (33.3 bar, blue symbols), 85 ppb (32.9 bar, green symbols), and 0.3 ppm $^3$He (33.3 bar, red symbols). Changes $\Delta\mu$ have been scaled by the values at the lowest temperature (18 mK) in order to compare temperature dependences. Open circles with lines are similarly scaled NCRI data from torsional oscillator measurements on 1 ppb $^3$He (31 bar, blue symbols[14]) and 0.3 ppm $^3$He (26 bar, red symbols[1]) samples.

**Supplementary Figure and Legend**

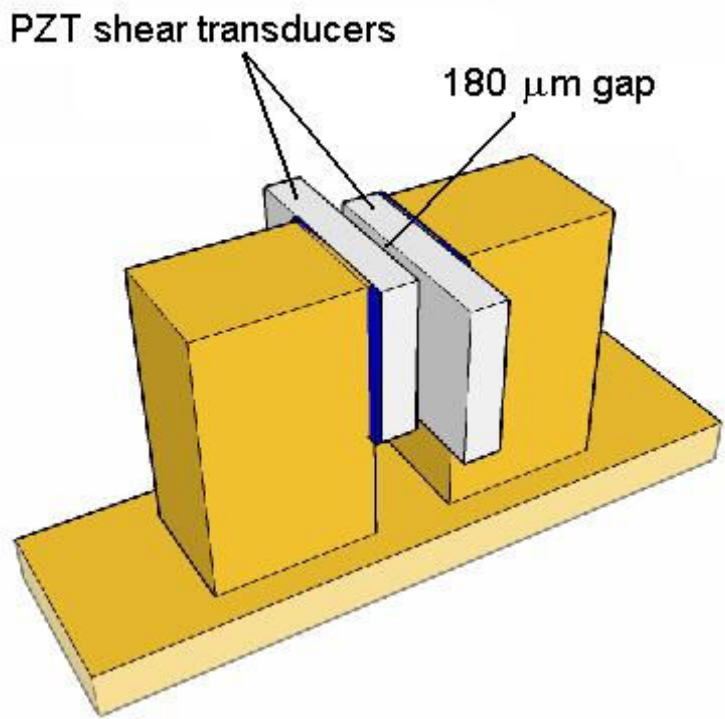

Supplementary Figure 1: Experimental setup for shear modulus and acoustic resonance measurements. Piezoelectric transducers (thickness 2.1 mm, width 9.6 mm, length 12.8 mm) are epoxied to brass backing pieces and rigidly mounted with their parallel faces separated by a small gap D ≈ 180 μm. Solid helium fills the gap where they overlap and surrounds the transducers.

**Supplementary Methods**

Helium was solidified using the blocked capillary method in a copper cell (volume ≈ 25 cm$^3$) which contained an *in situ* capacitive pressure gauge. Most crystals were grown from commercial $^4$He (nominal $^3$He concentration 0.3 ppm); we also made measurements with isotopically pure $^4$He (1 ppb $^3$He) and at intermediate concentrations produced by mixing the two gases. Supplemental Figure 1 shows the experimental arrangement. Displacements were generated and subsequent stresses were detected by two shear transducers[30] with a helium-filled gap (D ≈ 180 μm) between their faces. A voltage V applied to the driving transducer produces a shear displacement at its front face $\delta x = d_{15}V$ ($d_{15} = 0.12$ nm/volt for these transducers below 4 K). Below the resonance frequency of helium in the gap (~ 800 kHz), this creates a strain $\varepsilon_t = \delta x/D$, which then produces a shear stress $\sigma_t = \mu\varepsilon_t$ on the detecting transducer. The charge generated ($q = d_{15}A\sigma_t$, where A is the overlap area of the transducers) is detected as an output current $I = 2\pi f d_{15}A\sigma_t = 2\pi d_{15}^2(A/D)fV\mu$, where f is the drive frequency. The minimum detectable stress at 2000 Hz, set by the noise in our current preamplifier (~2.5 fA with 30 s averaging), is $\sigma_t$ ~ 2 x $10^{-5}$ Pa (corresponding to a displacement $\delta x$ ~ 2 x $10^{-16}$ m and strain $\varepsilon_t$ ~ $10^{-12}$). After subtracting a background due to electrical crosstalk (the signal with liquid $^4$He in the cell, a correction of less than 15%), the solid's shear modulus (~ 1.5 x $10^7$ Pa) is proportional to I/fV, which is nearly frequency independent below 4000 Hz. The transducers were offset horizontally by ~ 3 mm, providing exposed surfaces which could be used to excite and detect acoustic modes of solid helium outside the gap, surrounding the transducers. The first such resonance was at 7730 Hz at 0.5 K, roughly where predicted by an elastic model of our cell geometry.

**Supplementary Discussion**

Dislocations move in response to shear stress in their glide plane (related to the total stress by an orientation factor R which varies between 0 and 0.5). Their motion can be described by a "vibrating string" model[26] with line tension (~ $\mu b^2$) due to elastic energy and an effective mass per unit length (~ $\rho b^2$). External stresses provide a driving force $Rb\sigma_t$ and their motion is damped by thermal phonons. Below the dislocation loops' resonance frequencies (which are typically in the MHz range), inertia and damping are not important and they simply bow out between pinning points. Their displacement creates a strain which adds to the elastic strain and reduces the solid's shear modulus. In this regime the modulus change is independent of frequency: $\Delta\mu/\mu = -CR\Lambda L^2$, where C is a constant which depends on the distribution of loop lengths (~ 0.1 for a single length, and ~ 0.5 for an exponential distribution with average length L). In the absence of impurity pinning, L is the network length $L_N$, which is largest for low density dislocation networks with few intersections. In annealed crystals with well defined networks, $\Lambda L_N^2$ is a geometric constant (e.g., 3 for a cubic network) and $\Delta\mu$ can be nearly independent of the dislocation density. For the random orientations expected in polycrystalline samples, the average anisotropy factor is about 0.2, so dislocations can reduce $\mu$ by as much as 30%.

When impurities are added, the impurity pinning length $L_P$ can become smaller than $L_N$, reducing the dislocation strain. Impurity pinning is very effective, since a single pinning site at the middle of a loop reduces its contribution to m by a factor of four. At temperatures below $E_B$, impurities condense onto dislocations giving an enhanced concentration $x_D = x \cdot \exp(E_B/k_B T)$, where x is their bulk concentration. Pinning will be significant when $x_D$ increases to the point where a typical loop has an impurity bound to it (i.e., when $x_D \approx a/L_N$, where a is the atomic spacing along the dislocation). This implies that the shear modulus will recover to its intrinsic value below a pinning temperature which decreases with impurity concentration: $T_P \sim -(E_B/k_B) \cdot (\ln\{x \cdot L_N/a\})^{-1}$. Our results are consistent with this picture. Ultrasonic measurements[21,22] on helium single crystals gave dislocation densities $L \sim 10^6$ cm$^{-2}$ (polycrystals are expected to have higher densities). Typical loop lengths were $\sim 5$ $\mu$m, giving resonant frequencies $\sim 15$ MHz. Values of $R\Lambda L^2$ ranged from 1.0 in a low frequency measurement[19] to about 0.01 in the ultrasonic experiments. The dominant slip system for hcp $^4$He is edge dislocations gliding in the basal plane[31] and $^3$He impurities bind to these with $E_B/k_B$ in the range 0.7 K[19] to 0.3 K[27].

**Supplementary Notes**

30. PZT 51 supplied by Boston Piezo-Optics, Inc., 38 B Maple Street Bellingham, MA 02019 USA.

31. Tsuroka, F. and Hiki, Y. Ultrasonic attenuation and dislocation damping in helium crystals. *Phys. Rev.* **B20**, 2702 (1979).